%% file: main.tex
\documentclass[conference]{IEEEtran}

\usepackage{balance}  
\usepackage{graphicx} 
\usepackage{times}    
\usepackage{url}      
\usepackage[T1]{fontenc}
\usepackage{textcomp}
\usepackage[scaled=.92]{helvet} 
\usepackage{booktabs} 
\usepackage{ccicons}  
\usepackage{ragged2e} 
\usepackage{etoolbox}
\usepackage{xcolor}
\usepackage{xparse}
\usepackage{listings}
\usepackage{xspace}

\newcommand {\gustavo}[1]{{\color{blue}\bf{GS: #1}\normalfont}}

\newcommand {\bjoern}[1]{{\color{red}\bf{BH: #1}\normalfont}}
\newcommand {\andrew}[1]{{\color{purple}\bf{AH: #1}\normalfont}}

\newcommand{\tool}{TraceDiff}

\newtoggle{highlightchanges}
\togglefalse{highlightchanges}

\iftoggle{highlightchanges}{
   \newcommand {\changes}[1]{{\color{purple}{#1}\normalfont}}
}{
  \newcommand {\changes}[1]{{#1}}
}

\begin{document}
\lstset{language=Python,basicstyle=\small}
\title{\tool: Debugging Unexpected Code Behavior Using Trace Divergences}

\author{
\IEEEauthorblockN{
    Ryo Suzuki\IEEEauthorrefmark{1}, 
    Gustavo Soares\IEEEauthorrefmark{2}\IEEEauthorrefmark{3}\IEEEauthorrefmark{4}, 
    Andrew Head\IEEEauthorrefmark{3}, 
    Elena Glassman\IEEEauthorrefmark{3},\\ 
    Ruan Reis\IEEEauthorrefmark{2}, 
    Melina Mongiovi\IEEEauthorrefmark{2}, 
    Loris D'Antoni\IEEEauthorrefmark{5}, 
    Bj\"{o}rn Hartmann\IEEEauthorrefmark{3}
}
\IEEEauthorblockA{
    \IEEEauthorrefmark{1}University of Colorado Boulder,
    \IEEEauthorrefmark{2}UFCG,
    \IEEEauthorrefmark{3}UC Berkeley,
    \IEEEauthorrefmark{4}Microsoft Research,
    \IEEEauthorrefmark{5}University of Wisconsin-Madison\\
}
\IEEEauthorblockA{
    ryo.suzuki@colorado.edu, 
    gsoares@dsc.ufcg.edu.br, 
    \{andrewhead,eglassman,bjoern\}@berkeley.edu, \\
    ruanvictor@copin.ufcg.edu.br,
    melina@computacao.ufcg.edu.br,
    loris@cs.wisc.edu
}
\vspace{-0.9cm}
}


\maketitle

\input{0-abstract}
\input{1-introduction}

\input{2-related-work}
\input{3-formative-study}

\input{4-system}
\input{5-implementation}

\input{6-evaluation}

\input{7-discussion}

\input{8-acknowledgements}

\balance
\bibliographystyle{IEEEtran}
\bibliography{references}

\end{document}

%% file: 0-abstract.tex

\begin{abstract}
Recent advances in program synthesis offer means to automatically debug student submissions and generate personalized feedback 
in massive programming classrooms. When automatically generating feedback for programming assignments, a key challenge is designing pedagogically useful hints that are as effective as the manual feedback given by teachers. Through an analysis of teachers' hint-giving practices in 132 online Q\&A posts, we establish three design guidelines that an effective feedback design should follow. Based on these guidelines, we develop a feedback system that leverages both program synthesis and visualization techniques. 
Our system compares the dynamic code execution of both incorrect and fixed code and 
highlights how the error leads to a difference in behavior and where the incorrect code trace diverges from the expected solution. 
Results from our study suggest that our system enables students to detect and fix bugs 
that are not caught by students using another existing visual debugging tool.
\end{abstract}







%% file: 1-introduction.tex
\section{Introduction}

Personalized, timely feedback from teachers can help students get unstuck and correct their misconceptions~\cite{corbett2001locus, guo2015codeopticon}. 
However, personalized attention does not easily scale to massive programming classes~\cite{d2015can, glassman2015overcode}. In lieu of feedback, it is common for teachers in large classes to only provide test case suites, against which students can test their submissions. 

This substitution has some drawbacks. While a teacher might look at the student's submission and recommend reviewing a particularly relevant lesson or attempt to reteach an important concept, test case feedback can only point out how the student submission does not return the right answer. It can be difficult for a student to map failed test results back to a specific error in their code.

Recent advances in program synthesis promise to provide more specific personalized feedback at scale for programming assignments~\cite{head2017writing, kaleeswaran2016semi, rivers2015data, rolim2017learning, singh2013automated}.
These systems use program synthesis to learn code transformations that fix incorrect student submissions.
These transformations can then be turned into a hint sequence that begins with pointing hints (e.g., where the bug is) and ends with bottom-out hints (e.g., how to fix the bug)~\cite{rivers2015data, singh2013automated}.
For example, consider the following edit to fix an incorrect program:
\begin{lstlisting}
 def accumulate(combiner, base, n, term): 
-  if n == 1: 
+  if n == 0:
     return base 
  else: ...
\end{lstlisting}
This edit can be mapped to a series of hints such as pointing out the location of the bug (e.g., {\it ``Line 2 needs to be changed.''}) and suggesting the exact changes that the student needs to make (e.g., {\it ``In the expression {\tt if n == 1} in line 2, replace the value {\tt 1} with {\tt 0}.''}).

\begin{figure}[!t]
\includegraphics[width=3.5in]{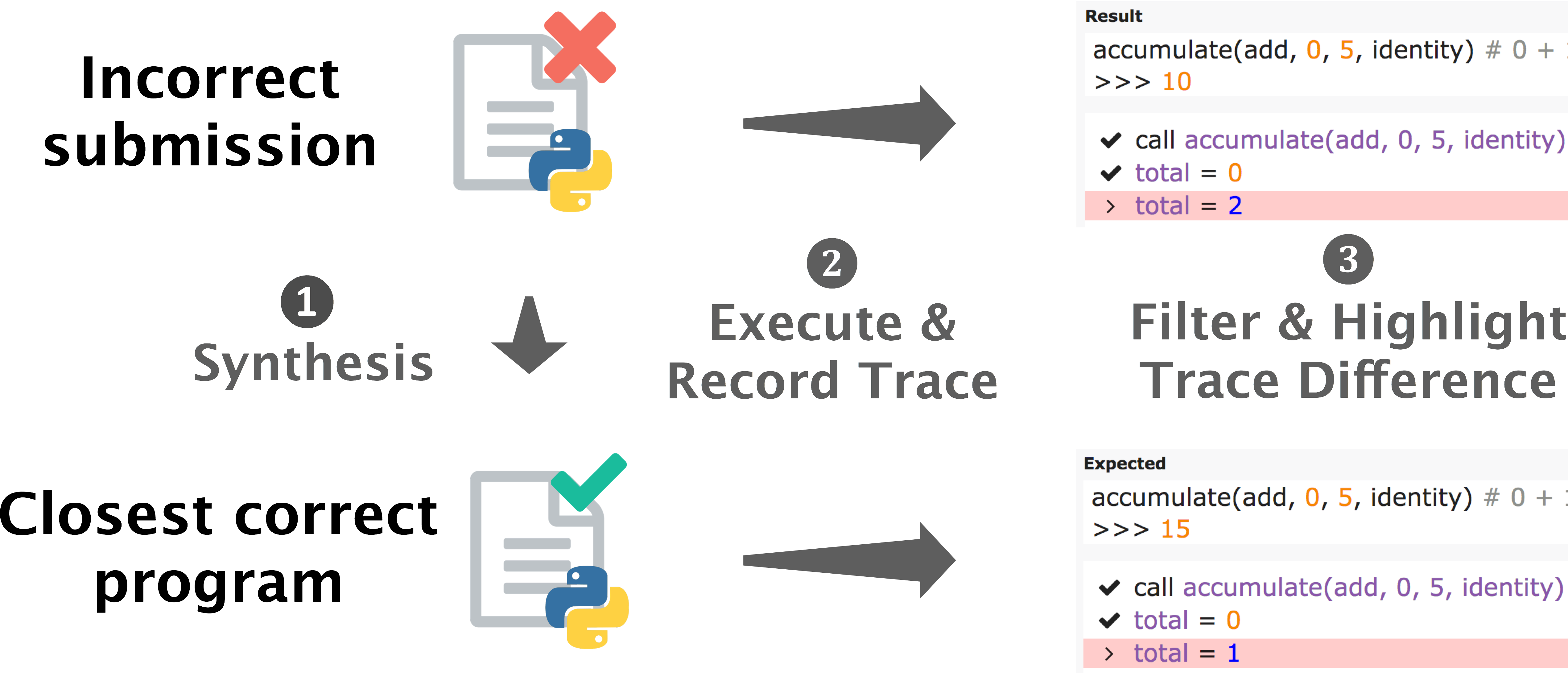}
\centering
\caption{Our system finds the closest correct program to an incorrect student program, executes both correct and incorrect programs, and shows where the traces of the two programs begin to diverge.}
~\label{fig:system-concept}
\vspace{-0.5cm}
\end{figure}

However, appropriate strategies for turning fixes into more pedagogically useful programming feedback remain an open research problem.
Well-designed feedback not only instructs how to fix the bug, but also facilitates conceptual understanding of the underlying cause of the problem~\cite{shute2008focus, vanlehn2006behavior}.
In our observations, 80\% of the time, teachers employ at least one the following hint-giving strategies: reminding students of relevant resources, explaining incorrect state or behavior, and diagnosing the cause of the problem.
Automatic pointing and bottom-out hints do not reflect these hint-giving strategies.
Although some of these hint-giving strategies may be difficult to automate, we hypothesize that automatic synthesized feedback can be improved such that it provides high-level hints that explain why the student-written code is wrong and how the synthesized code fix affects code behavior~\cite{suzuki2017exploring}.

In this paper, we present \tool{}, an automatic feedback system
that leverages both program synthesis and  program visualization techniques to provide interactive personalized hints for introductory programming assignments (see Figure~\ref{fig:system-concept}).
Given the student's incorrect code and a synthesized code fix, our system first performs dynamic program analysis to capture the execution of both the incorrect and fixed code, comparing internal states and runtime behaviors.
Then, the system highlights how and where the trace of the incorrect code diverges from the trace of the fixed code.
To enable users to interactively explore the behavior of the code, we incorporate Python Tutor~\cite{guo2013online}, an existing visual debugging tool frequently used in introductory classes, into our interface.
Our system is based on the following strategies:
\begin{enumerate}
\item {\bf Highlight behavior that diverges from the nearest solution:} It compares the executions of incorrect and fixed code and highlights the point when their control flows diverge. 
\item {\bf Focus attention by extracting important steps:} It filters the execution traces which are only relevant to the student's mistake to focus attention.
\item {\bf Explore behavior through interactive program visualization:} It integrates Python Tutor to allow interactive exploration of collected code traces. 
\item {\bf Map erroneous concrete values to their cause by abstracting expressions:} It enables the student to interactively map a concrete value (e.g., {\tt sum = 3} and {\tt return 11}) back to the expressions that computed these values, such as variables and function calls (e.g., {\tt sum = add(1, 2)} and {\tt return total }) to help locate the cause of a test case failure. 
\end{enumerate}

Our system design is informed by a formative study where we analyzed 132 Q\&A posts from a discussion forum from an introductory programming class. We also interviewed a teaching assistant from this class.
Based on the strategies the TAs employ to answer student questions, we first identified three high-level design guidelines that effective feedback systems should follow: 
(1) encourage students to explore code execution with a visual debugger,
(2) describe how actual behavior differs from expected behavior,
(3) refer to concrete code locations and behavior to provide a starting point for exploration.
We then investigated how these strategies could be automated in a feedback interface.
Based on these guidelines, we designed the features of a system and integrated these features with an interactive debugging interface.

To evaluate if our system enables a more efficient debugging experience than current interactive debugging tools, we conducted a controlled experiment with 17 students where participants were asked to debug incorrect student code from introductory Python programming assignments and compare \tool{} with the Python Tutor interface.
During a 60-minute session, each participant was asked to perform two bug-fixing tasks for each incorrect code: (1) locate the bug and (2) fix the bug. 
We evaluated whether or not each participant correctly answered these questions and measured the time spent to complete these tasks.
The result shows that, for one of the incorrect code, only participants using \tool{} were able to fix it (5 out of 9), while none of the 8 participants using Python Tutor could fix it.
Although we have not found statistically significant differences in the quantitative measures of the two groups, 64.7\% of the participants believed that \tool{} was the more valuable to identify and fix the bugs and 29.4\% thought that both tools were equally important (only 5.9\% preferred Python Tutor). 
In response to 7-point Likert scale questions, participants significantly preferred TraceDiff over Python Tutor in four out of five dimensions: overall usefulness, usefulness to identify, understand, and fix the bugs.


In summary, in this work, we contribute: 
\begin{itemize}
\item a characterization of key design guidelines for effective programming feedback that can be generated by state-of-the-art synthesis techniques, informed by a formative study.
\item the implementation of hints in an interactive debugging interface, appropriate for deployment and evaluation in a massive programming classroom.
\item quantitative and qualitative results of a controlled experiment with 17 students where we compare \tool{} with Python Tutor interface.
\end{itemize}

%% file: 2-related-work.tex
\section{Related Work}

\subsection{Automated Feedback for Programming Assignments}

Intelligent tutoring systems (ITSs) often supply a sequence of hints that descend from high-level pointers down to specific, bottom-out hints that spell out exactly how to generate the correct solution. For example, in the Andes Physics Tutoring System, hints were delivered in a sequence: \textit{pointing}, \textit{teaching}, and \textit{bottom-out}~\cite{Vanlehn2005}. 
ITSs have been historically expensive and time-consuming to build because they rely heavily on experts to construct hints. 

Recently, researchers have demonstrated how program synthesis can generate some of the personalized and automatic feedback typically found in ITSs (e.g.,~\cite{kaleeswaran2016semi, rivers2015data, rolim2017learning, singh2013automated}).
For example, AutoGrader~\cite{singh2013automated} can identify and fix a bug in an incorrect code submission, and then automatically generate sequences of increasingly specific hints about where the bug is and what a student needs to change to fix it. 

High-level hints that point to relevant class materials or attempt to reteach a concept can be difficult to automatically generate because they require more context or the deep domain knowledge of a teacher.
To leverage the teacher's high-level feedback at scale, CodeOpticon~\cite{guo2015codeopticon} provides a tutoring interface that helps teachers provide synchronous feedback for multiple students at once.
Recent work has also demonstrated how program analysis and synthesis can be used as an aid for a teacher to scale feedback grounded in their deep domain knowledge~\cite{glassman2015overcode, head2017writing}.
While scaling the return on teacher effort, these systems still require teachers to manually review and write hints for incorrect student work. 

In contrast to prior work on scaling up teacher-written feedback, this paper focuses on \textit{fully automated} approaches to provide high-level hints, specifically for the context of writing code.
D'Antoni et al. have explored the similar design challenge of automatically generated hints for the domain of finite automata~\cite{d2015can}.
Taking inspiration from this work, we aim to generate high-level hints in the domain of introductory programming assignments. 


\subsection{Design of Interactive Debugging Tools}

One of the major challenges in learning to program is to relate code to the dynamics of program execution~\cite{du1986some}.
In an introductory programming course, many novice students have difficulties and misconceptions due to a lack of understanding of dynamic program execution~\cite{ragonis2005understanding}.
One practical way to alleviate this cognitive difficulty is to visualize execution. 
Recently, researchers have proposed many program visualization tools (see ~\cite{sorva2012visual} for a comprehensive review).
These tools typically execute the program, store a snapshot of internal states at each execution step, and show a visual representation of runtime states such as stack frames, heap objects, and data structures~\cite{guo2013online}.
Recent studies have found that using these program visualization tools can be pedagogically effective if students actively engage with the tool~\cite{hundhausen2002meta, sorva2013review}.

However, as program complexity increases, such visualizations can become confusing~\cite{urquiza2004survey}, and navigating the traces may become-time consuming. 
Alternatively, recent debugging interfaces like Whyline~\cite{ko2008debugging} and Theseus~\cite{lieber2014addressing} provide an overview of execution behavior and let a user find the cause of a bug through interactive question-answering or retroactive logging. 
Inspired by this prior work, this paper aims to augment program visualization to enable more efficient review of program traces.
One design challenge is how to focus a student's attention on the differences between what the code does and what it is expected to do. 
Previous work has demonstrated that important divergent runtime behavior can be detected and highlighted in the domain of web application debugging~\cite{burg2013interactive, burg2015explaining, oney2009firecrystal}.
To apply similar design insights debugging programming assignments, we leverage a program synthesis technique~\cite{rolim2017learning} that identifies potential corrections to programming assignments; with a corrected version of code, we extract traces that highlight the difference in the code's current and expected behavior.

%% file: 3-formative-study.tex
\NewDocumentEnvironment{postquote}{m}
    {``\itshape\kern-.05em}
    {\normalfont\kern-.05em'' (post #1)}

\section{Formative Study}

To understand the current limitations of automatic hint delivery and opportunities to improve it, we observed the hint-giving practices of teachers in a local introductory CS course as they helped students debug incorrect code for programming assignments.  
We analyzed 132 Q\&A posts from the CS course's online discussion forum where instructors answered students' debugging questions.
Additionally, we conducted a semi-structured interview with a teaching assistant from the same course to gain insight into the patterns of hint-giving that we observed in the online discussions.
This analysis yielded three design guidelines that motivated the design of \tool{}'s hint-giving affordances.


\subsection{Procedure}

We collected 132 posts from the discussion board that pertained to one assignment from a recent course offering.
One author performed open coding on all posts, eliciting common themes in the structure and content of teachers' hints.
Two authors reviewed the themes together, refining the themes into types of hints, and deciding on definitions and concrete examples for each of type of hint.
The two authors performed axial coding independently, tagging the types of hints they observed in each teacher response.
These two authors resolved all discrepancies in tagging results by reviewing each tag until they reached consensus.
Then, the two authors identified three design guidelines by reviewing high-level common strategies that share among the themes.
After that, we had a 30-minute semi-structured interview where we asked what kind of questions the students frequently ask to the TAs, how TAs answer these questions, and what the high-level strategy is for appropriate feedback.

\subsection{Design Guidelines}


\changes{
Based on the observation and analysis of teachers' hint-giving strategies, we describe three design guidelines, which motivated our interface design of \tool{}.
}

\subsubsection*{D1: Encourage students to explore code execution with a visual debugger}

For novice learners, it is difficult to understand complex code executions without visual aids.
Program visualization tools help with this problem:
we observed one of the most common feedback (19 times) was to tell students to run their code in Python Tutor~\cite{guo2013online}, an interactive code visualization tool.
Hence the decision of integration the other hint strategies with the Python Tutor interface.

\subsubsection*{D2: Describe how actual behavior differs from expected behavior}

Well-designed feedback facilitates productive debugging by illustrating the relationship between the symptoms and the cause of the error~\cite{ko2008debugging, ko2004six}.
TAs often diagnose the student error through abstracting the suggestion or providing the high-level description of the cause of error (19 times).
\begin{postquote}{60}
``Runtime Error - Maximum recursion depth exceeded in comparison'' message means that you do not have a base case that can stop the program from running your recursive calls
\end{postquote}.
Although it can be difficult to automate hints that provide a conceptual description of the student error, TAs diagnose
these kinds of errors by comparing what the code does with what it is expected to do.
\begin{postquote}{82}
In your \texttt{n == 1} base case, you should be calling \texttt{term(1)} and not \texttt{term(0)}. Remember that the \texttt{term} function is only defined from \texttt{1} to \texttt{n} inclusive
\end{postquote}.
Hence the decision to provide feedback by highlighting behavior that diverges from the closest solution obtained using a program synthesis back-end.


\begin{figure*}[!t]
\includegraphics[width=7.0in]{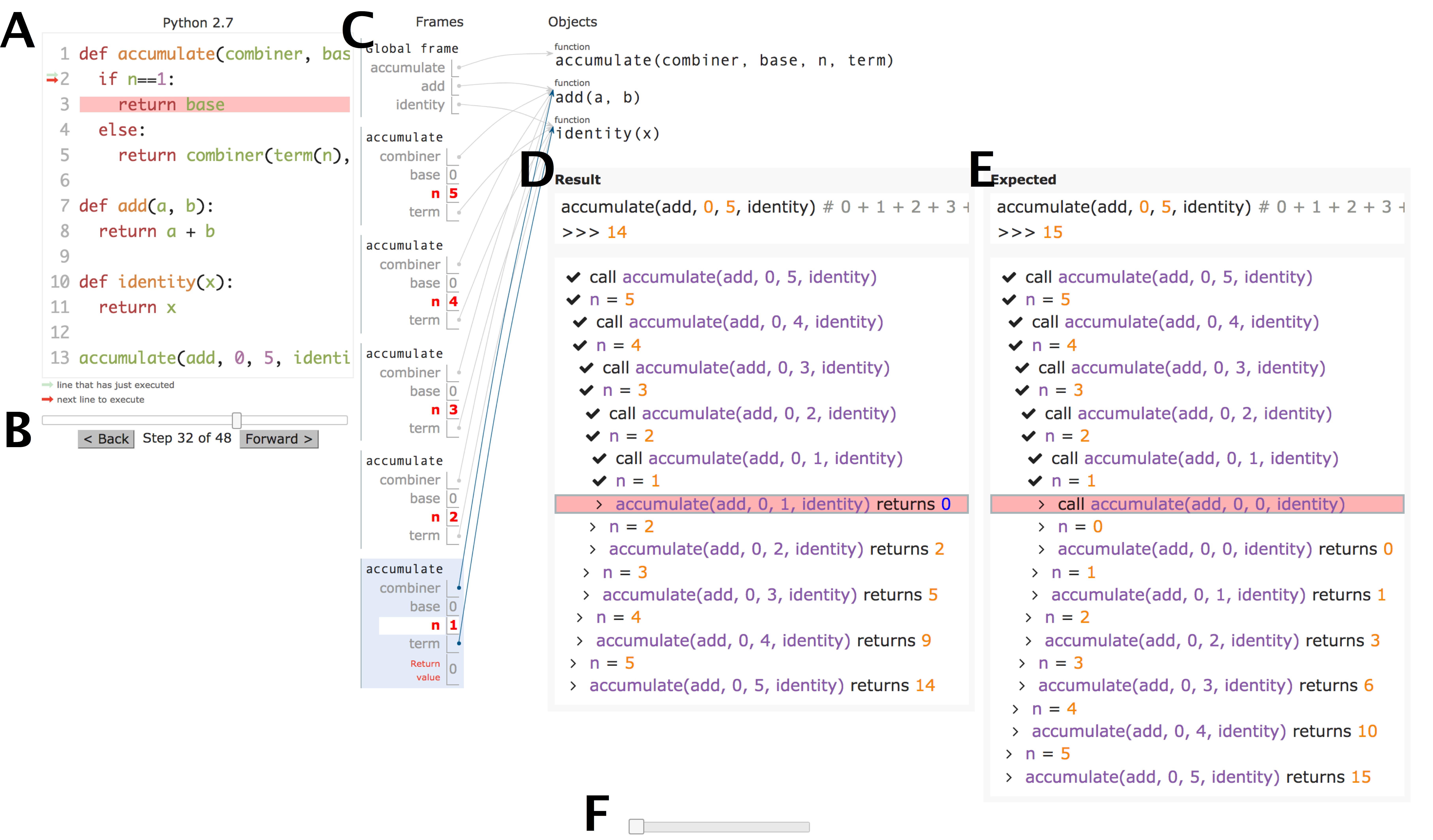}
\centering
\caption{The system interface consists of a code editor (A), a visual debugger (B-C), extracted code traces (D-E), and a slider for interactive abstraction (F).}
~\label{fig:system-overview}
\vspace{-0.5cm}
\end{figure*}

\subsubsection*{D3: Refer to concrete code locations and behavior to provide a starting point for exploration}

As the program becomes large and complex, a visual code execution can be confusing and difficult to track ~\cite{urquiza2004survey}.
To give students focus, some TAs pointed to specific locations or provided scaffolding questions (3 times) along with suggesting Python Tutor.
\begin{postquote}{74}
Try to examine the code in Python Tutor. What happens when you call accumulate? Is the combiner that you're passing on to accumulate making a decision based on the predicate for every number in the sequence?
\end{postquote}

In addition to a position in code, TAs also pointed out a particular code structure or a specific point in the behavior:
\begin{postquote}{50}
Look carefully at your \texttt{else} case and also the condition of your \texttt{if}. Is it doing what you expect it to? 
\end{postquote}
TAs provided data and behavior hints to help illuminate why a student's code fails.
\begin{postquote}{28}
Think about what the counter value is, and what the total value is. Is this correct? Remember, ping pong looks like: \\
\texttt{total = 1 2 3 4 5 6 [7] 6 5} \\
\texttt{count = 1 2 3 4 5 6 7 8 9} \\
for the first 9 elements.
\end{postquote}
Therefore, we designed our hint interface to give a quick overview of code execution that focuses attention, while allowing to explore the detail of the trace with a visual debugger.

%% file: 4-system.tex
\section{Interface Design}

When a student submits an incorrect program to the \tool{} system, the system back-end synthesizes a fix that corrects the program. This fixed program will be both correct and syntactically close to the student's incorrect submission. The interface can leverage this pair of incorrect and correct programs to show the student the difference between the actual behavior of their submission and expected behavior which would pass all the test cases for the assignment.

The system executes the incorrect and fixed programs, and stores a snapshot of both their internal states at every execution point.
Using this information, the system interface, shown in Figure~\ref{fig:system-overview}, renders execution traces of both the incorrect (D) and fixed (E) programs side-by-side.
To help the student find the behavioral differences, the interface highlights where the incorrect program diverges from the fixed one, both in the original code (Figure~\ref{fig:system-overview}~A) and the trace (Figure~\ref{fig:system-overview}~D,~E). The student can rapidly scan over the causes of the unexpected behavior by scrubbing the slider (Figure~\ref{fig:system-overview}~F) below the traces.
The student can inspect these behavioral differences further by clicking on an item in the trace, triggering the Python Tutor interactive visualization interface to render the stack frames and objects at that point of execution (Figure~\ref{fig:system-overview}~C) and indicate which line of code was just executed (Figure~\ref{fig:system-overview}~B).

We design our interface \changes{by building multiple prototypes through a human-centered design process} to address all three design goals identified in our formative study.
The main features of our interface are summarized below:
\begin{enumerate}

\item {\bf Filter}: extracts important control flow steps by identifying a list of variables and function calls that take on different values to focus students' attention (D3) 
\item {\bf Highlight}: compares the execution of incorrect and fixed programs and highlights a point where the control flow diverges to help identify the bug (D2, D3) 
\item {\bf Explore} integrates the Python Tutor to enhance the exploration of collected code traces (D1) 
\item {\bf Abstract}: enables the student to interactively map a concrete value back to the expressions that computed it and helps locate the cause of the bug (D3) 
\end{enumerate}


\subsection{Filter}
\begin{figure}[!h]
\includegraphics[width=3.5in]{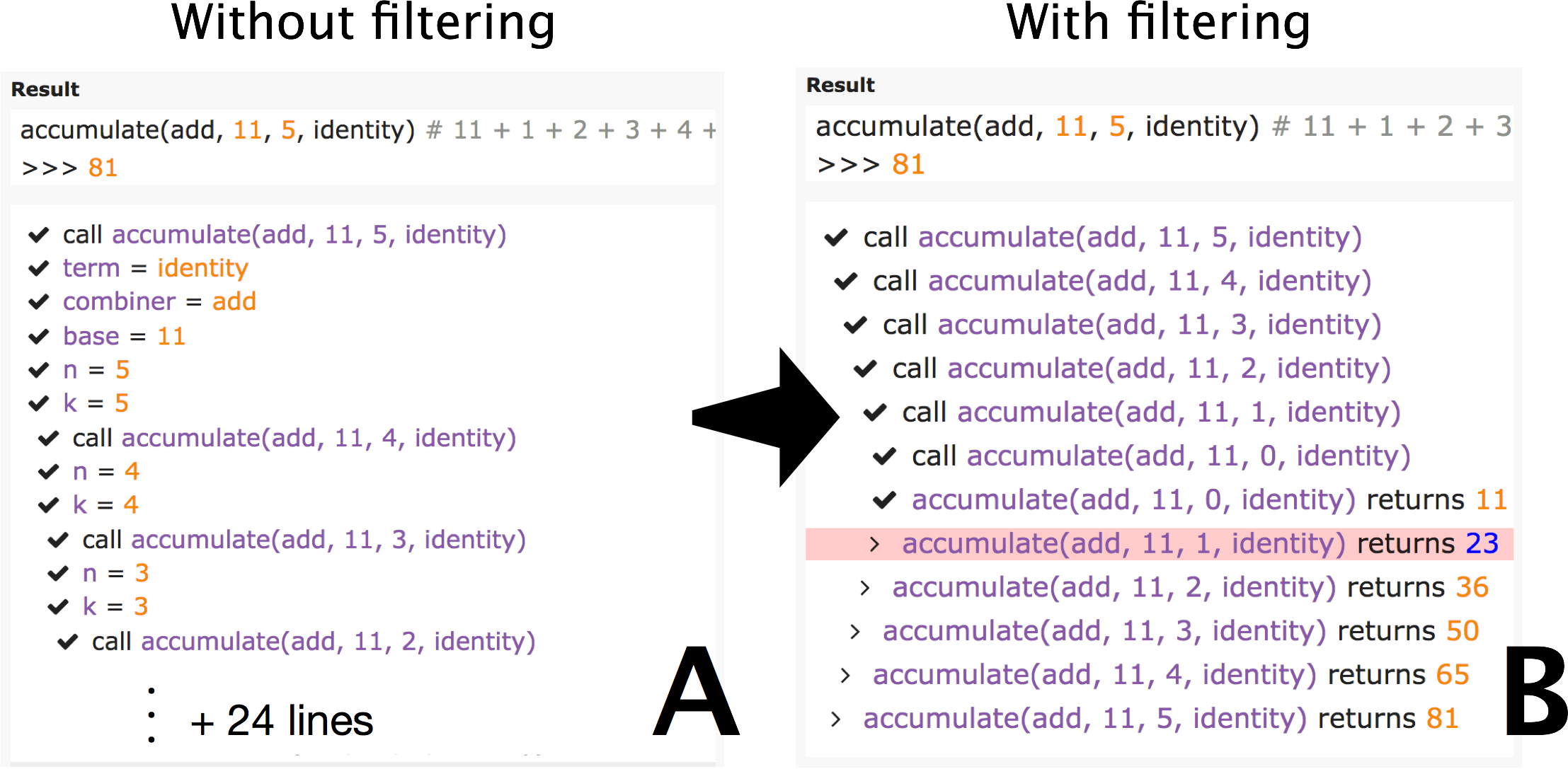}
\centering
\caption{Filter function: The system back-end synthesizes a fixed program and generates execution traces for both the student's incorrect program and the fixed program. The system extracts the steps in the traces for only those variables, calls, and returns that eventually diverge from the fixed program. (A) illustrates the code traces without filtering and (B) shows the filtered execution traces of the same programs.}
~\label{fig:system-filter}
\vspace{-0.2cm}
\end{figure}

The system shows execution traces of both the incorrect and fixed programs side-by-side (Figure~\ref{fig:system-overview}~D,~E).
However, as the execution flow becomes complex, students can be overwhelmed with too much trace information \changes{and find it difficult to grasp an overview of the behavior}.
We filter the execution trace to better focus the student's attention on potentially relevant steps (D3). 
For all variables, function calls, and return statements, the system compares the sequence of assigned, passed, or returned values between the two programs throughout their traces. All variables, calls, and returns with equivalent sequences are filtered out, since their behaviors are identical. Both matching and divergent assignments of values for the remaining variables are shown in the interface. Figure~\ref{fig:system-filter} illustrates an example of the filtering feature.

%

\subsection{Highlight}
\begin{figure}[!h]
\includegraphics[width=3.5in]{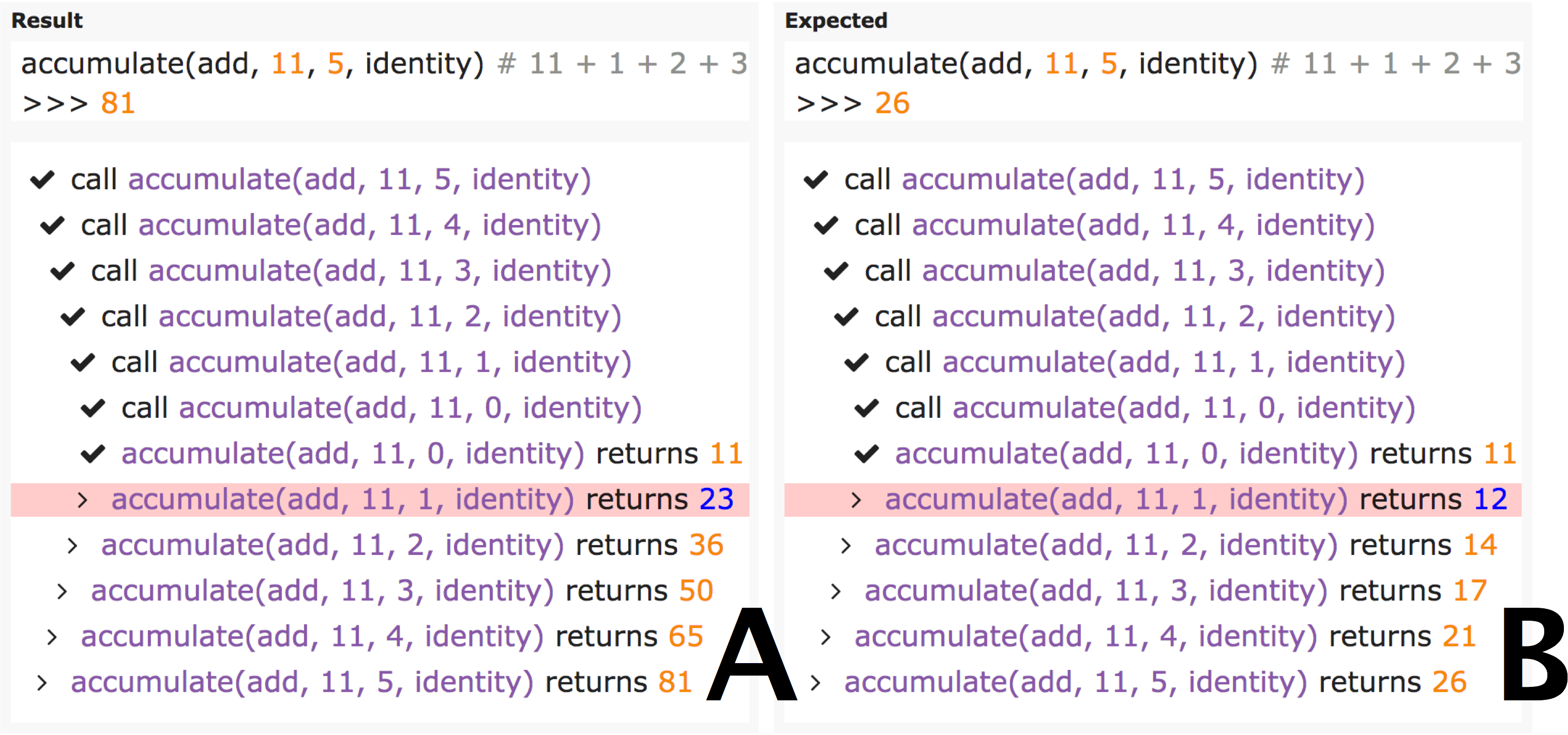}
\centering
\caption{Highlight function: Once the system filters the execution traces, it compares the execution of the incorrect program with the fixed program and highlights the step where the incorrect control flow diverges from the fixed control flow.}
~\label{fig:system-highlight}
\vspace{-0.2cm}
\end{figure}

The filtered traces show both consistent and inconsistent steps for the incorrect and fixed programs.
To help students locate the error in their program, we identify the first step where the values \changes{(e.g., values of variables, function calls, and return statements)} diverge between incorrect and fixed programs and highlight it in the interface (see Figure~\ref{fig:system-highlight}).

Figure~\ref{fig:system-highlight} shows the comparison of the actual and expected behavior of the same incorrect submission shown in Figure~\ref{fig:system-filter}. 
One can quickly see that the difference in the test result originates from the {\tt accumulate} function returning 11, 23, 36, 50, 65, 81 instead of 11, 12, 14, 17, 21, 26.
Thus, this comparison can address guideline D2 by scaffolding the understanding of the cause of the error.

By comparing the execution history, our system can also help a student locate the error (D3). 
When the system detects that a variable value in an incorrect program diverges from its corresponding variable value in the fixed program, the system can highlight the difference between the actual and expected values. 
For example, Figure~\ref{fig:system-highlight} also indicates that the execution traces diverge when {\tt accumulate(add, 11, 1, identity)} returns 23 instead of 12, so the error of the code should be associated with this line. 
To help students relate a step in the trace with a position in the code, the system highlights the relevant line number in the code when the user hovers over a step. 
In contrast to existing feedback, this helps not only to identify the location of the bug, but also to understand why the code fails by suggesting how this difference at a particular point affects the return value.

\begin{figure}[!h]
\includegraphics[width=3.5in]{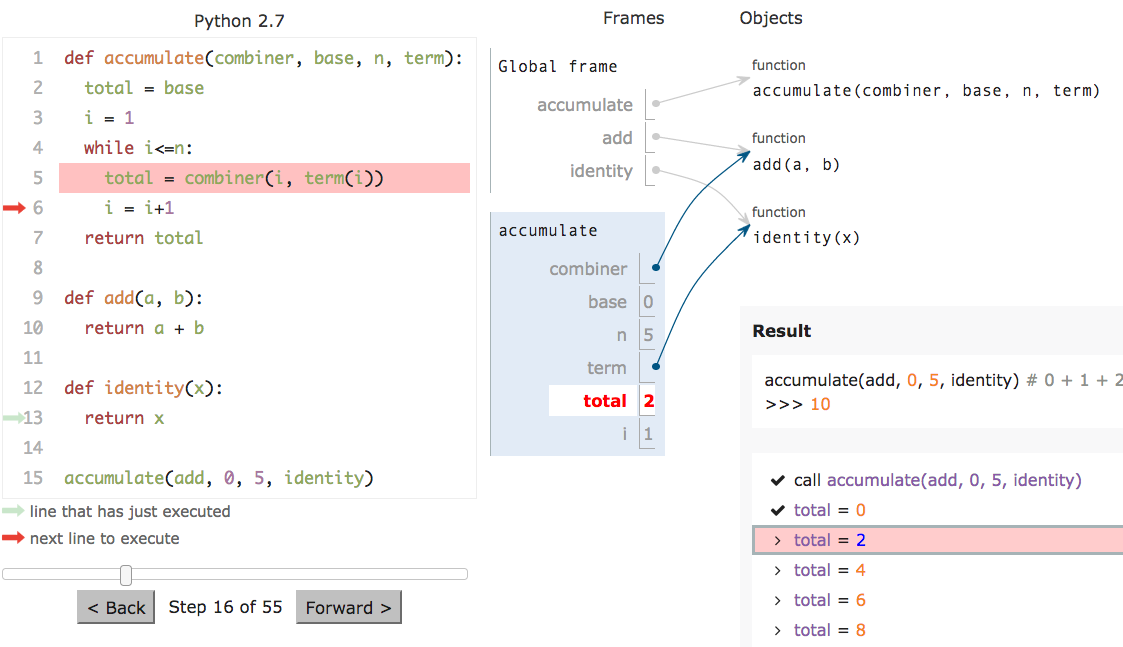}
\centering
\caption{Explore function: When the student hovers over the execution trace, the system highlights the line of code (e.g., line 5) responsible for the current step of the execution. By clicking a step in the execution trace panel (e.g., the highlighted step at total = 2), the system shows the corresponding code execution with the Python Tutor visual interface. Within the visual stack frame, the system also highlights the variables and function calls that take different values between the incorrect and correct programs.}
~\label{fig:system-explore}
\vspace{-0.2cm}
\end{figure}

\subsection{Explore}

Once a student locates a bug, \tool{} helps the student understand the unexpected program behavior with visual code execution (D1, see Figure~\ref{fig:system-explore}).
\changes{
While the filtering function can provide an overview of the divergence, it may reduce the context of the code traces.
Thus, to allow the student to access the full context of the change,
}
we integrated the Python Tutor~\cite{guo2013online} visual debugging tool, using brushing and linking to connect execution traces with Python Tutor's code editor and debugger.  
By clicking a specific function call in the execution trace, the system jumps to the relevant step and visualizes the program state at that step using the Python Tutor visual debugging tool.
\tool{} highlights variables in the Python Tutor visualization that differ between the actual and expected program behavior in red.

\subsection{Abstract}
While the highlighting feature helps compare low-level (concrete) data and behavior between the correct and incorrect programs, it does not help link between concrete differences in values and differences in high-level (abstract) structure of the code. 
One possible way to address this issue is to visualize both the incorrect and fixed programs with the Python Tutor interface. 
However, we found that comparing the code execution with multiple diagrams was difficult to comprehend, particularly when the control flow is significantly different.
Instead, we design an alternative interface inspired by \changes{Bret Victor's learnable programming~\cite{victor2012learnable} and } a ladder of abstraction~\cite{victor2011up}. 

\begin{figure}[!h]
\includegraphics[width=3.5in]{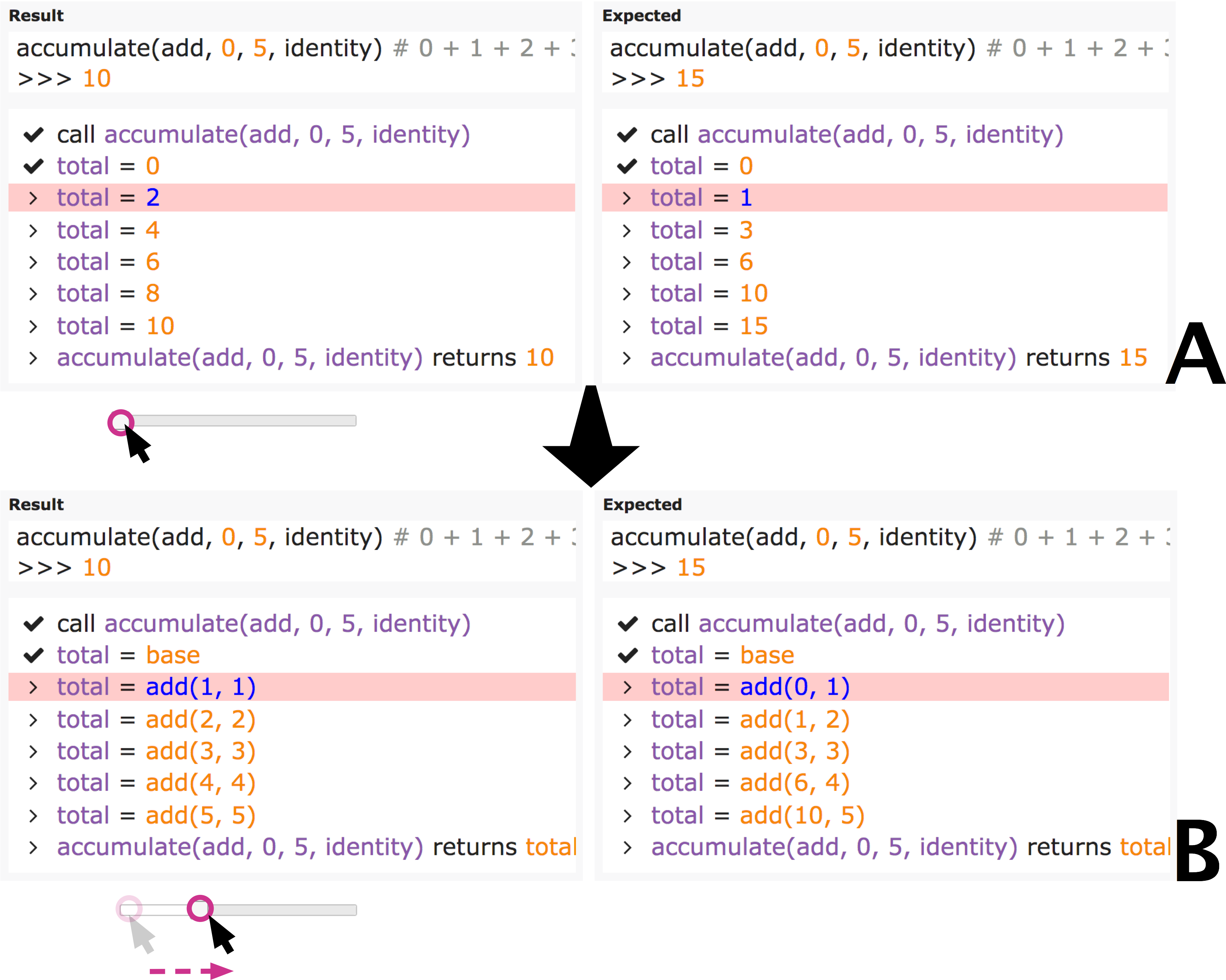}
\centering
\caption{Abstract function: By scrubbing the slider, the system converts the concrete value into the abstract expression that shows how the value is computed. It enables the student to rapidly scan over the causes of the unexpected behavior by comparing differences in code expressions.}
~\label{fig:system-abstract}
\vspace{-0.2cm}
\end{figure}

Figure~\ref{fig:system-abstract} A shows the behavior of the following incorrect code:
\begin{lstlisting}
def accumulate(combiner, base, n, term):
  total = base
  i = 1
  while i<=n:
    total = combiner(i, term(i))
    i = i+1
  return total    
\end{lstlisting}
This program is almost correct, except the first argument of the {\tt combiner} call is wrong. To pass all the test cases, the expression {\tt combiner(i, term(i))} on line number 5 can be changed to {\tt combiner(total, term(i))} to correctly accumulate the value.
The difference in concrete values assigned to {\tt total} that are returned by the {\tt combiner} function call, shown in Figure~\ref{fig:system-abstract}~A, does not reveal the incorrect argument in the {\tt combiner} function call.
To help students map the difference in concrete values to the incorrect expressions in the code, students can scrub the slider to see the expressions responsible for computing differing concrete values (Figure~\ref{fig:system-abstract}~B). 
The information the student can see as a result of that scrubbing indicates that the second argument of the {\tt add} function is correct, but the first argument is different.
This feature helps uncover the cause of the error, particularly for complex mistakes (D3). 

%% file: 5-implementation.tex
\section{Implementation}

The front-end of \tool{} is written in JavaScript and depends on D3.js for data-binding and drawing the user interface.
Our back-end comprises four components that perform code transformation synthesis, execution trace collection, trace differencing, and value abstraction.
\changes{
The source code of both front-end and back-end of \tool{} is open source and available on GitHub\footnote{https://github.com/ryosuzuki/trace-diff}.
}

\subsection{Code Transformation Synthesis}

Code transformations enable \tool{} to correct a student's incorrect code so the behavior of the incorrect and corrected code can be compared.  To learn these transformations, we leverage Refazer~\cite{rolim2017learning}, a tool 
that, given pairs of incorrect and corrected code, learns general, code-correcting abstract syntax tree transformations.
\changes{Refazer represents these transformations in a Domain-Specific Language (DSL). It applies algorithms that deduce the set of transformations that can be generated from the code examples and algorithms that rank these transformations so that the top ones have higher probability to be the ones that will be successfully applied to other programs.}

For each programming assignment, we extracted pairs of incorrect and corrected code by mining student submission histories to a course autograder.  Each pair contains a correct submission according to the test cases and an incorrect submission, which is the first submission before the correct submission.

Once we have learned a set of code transformations for an assignment, 
we fix incorrect code by applying code transformations to incorrect submissions one-by-one.  When a code transformation can be successfully applied and causes the code to pass all assignment test cases, the code is marked as corrected.  We save this corrected copy of the submission, so that its execution trace can be compared to the trace of the original incorrect submission.

\subsection{Execution Trace Collection}

We use the Python debugging module \texttt{pdb}~\cite{pdb} to collect execution traces of both the incorrect and corrected code.  We step through both versions of the code separately, serializing the program state at each step.  The program state comprises the names of local and global variables, values of those variables, the stack, and the line number of the currently executing code.  This information allows us to later compare sequences of values each variable from each program's execution.


\subsection{Trace Differencing}
\tool{} extracts the key events from the collected traces by filtering out variables and calls that do not lead to a divergence between incorrect and fixed programs. 
Since 
the sequences of correct and incorrect code execution may have  different control flow or different line numbers,
\tool{}  performs the comparison on the sequence of updates of each variable value and return value of a function, and detects value updates which differ at the same relative index in their sequence.
For example, the code shown in Figure~\ref{fig:system-explore} has six variables (\texttt{combiner, base, n, term, total,} and {\tt i}). 
Among these variables, only \texttt{total} has  sequences of updated values that diverge (e.g. \texttt{[0, 2, 4, 6, 8, 10]} and \texttt{[0, 1, 3, 6, 10, 15]}). The other variables have the same update sequences (e.g., i is the same sequence of updates \texttt{[1, 2, 3, 4, 5, 6]} for both incorrect and fixed code trace).
This technique allows us to extract only traces for the key variables the student should pay attention to.

\begin{figure}[t]
\includegraphics[width=3.2in]{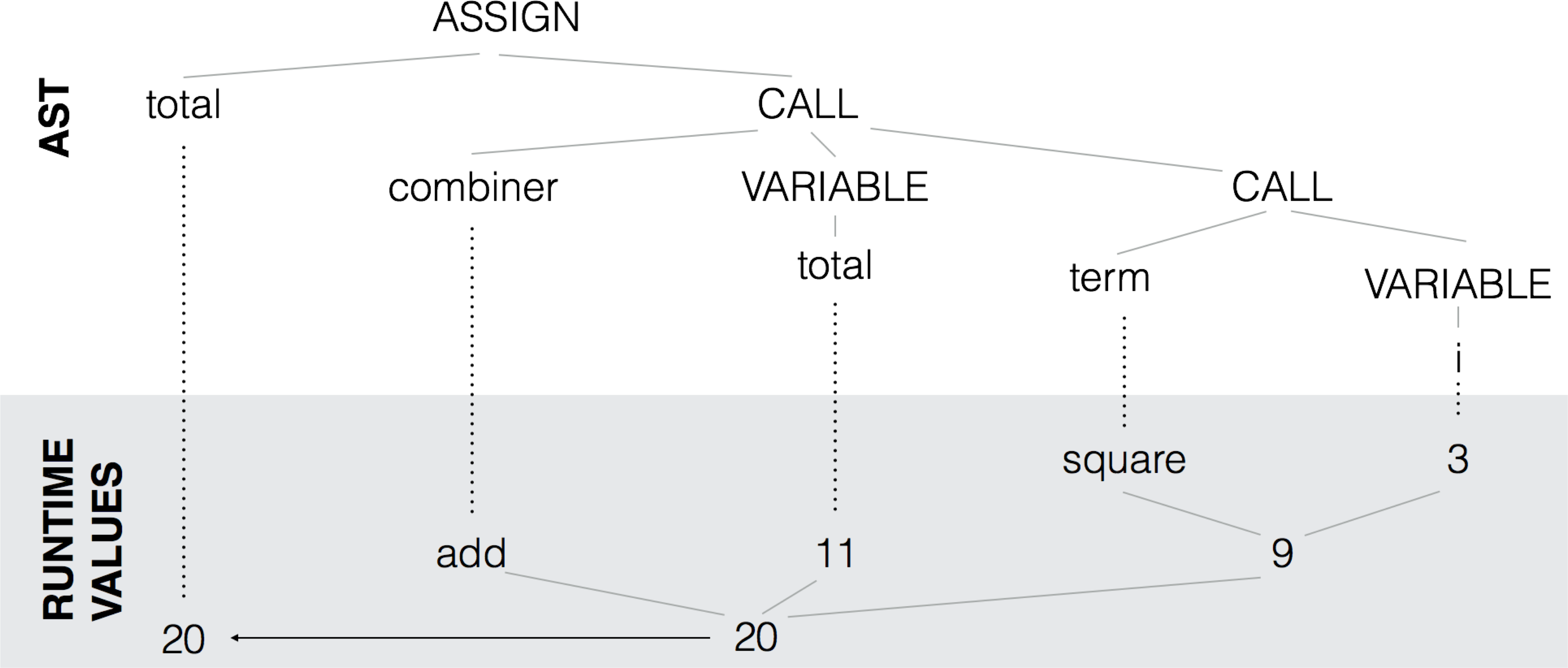}
\centering
\caption{Example AST for the statement \texttt{ total = combiner(total, term(i))} on top; dynamic values observed at runtime below.}
~\label{fig:ast}
\vspace{-0.2cm}
\end{figure}

\subsection{Abstracting Values into Expressions}
To abstract the observed values back into expression form, we leverage a combination of dynamic and static program analysis techniques.
Given the line of code, we first parse the line into its Abstract Syntax Tree (AST). 
For example, the code in Figure~\ref{fig:system-explore} at line 5 {\tt total = combiner(i, term(i)) 
} can be parsed into the tree shown in Figure ~\ref{fig:ast}, top.

We recursively evaluate each AST node with the current state obtained from the dynamic program analysis.
For example, when the internal states are {\tt combiner = add, term = identity, i = 2, total = 4}, we can construct the matching {\em value tree} 
shown in Figure ~\ref{fig:ast}, bottom.
We can traverse this tree from the single output value back up to intermediate values and back to corresponding AST nodes to generate abstract expressions.
Traversing this tree  generates the following sequence of increasingly abstract expressions:
\texttt{4}
$\rightarrow$
\texttt{add(2, 2)}
$\rightarrow$
\texttt{combiner(2, 2)}
$\rightarrow$
\texttt{combiner(2, identity(2))}
$\rightarrow$
\texttt{combiner(2, term(2))}
$\rightarrow$
\texttt{combiner(2, term(i))}
$\rightarrow$
\texttt{combiner(i, term(i))}

\changes{
All concrete values in the run-time execution can be abstracted until we get the root node of the tree (e.g., {\tt total = combiner(i, term(i))} in Figure 7). 
However, if the system allows students to see all abstraction steps, it may reveal the underlying actual code. 
Therefore, we heuristically set the three-level of abstraction as the stopping point to avoid revealing a bottom-out hint.
}

%% file: 6-evaluation.tex
\section{Evaluation}

To see if \tool{} can help students debug their code efficiently, we conducted a controlled experiment and evaluated our interface alongside the Online Python Tutor interface.

\subsection{Research Questions}
\begin{itemize}
\item {\bf RQ1:} Can \tool{} help students identify and fix more bugs than when just using Python Tutor?
\item {\bf RQ2:} Can \tool{} help students fix bugs faster than when just using Python Tutor? 
\item {\bf RQ3:} Which tool, \tool{} or Python Tutor, do students perceive to be more useful for fixing bugs?
\end{itemize}

\subsection{Method}
We recruited 17 students (male: 15, female: 2; undergraduate: 13, graduate: 4) from a local university to participate in this study. 
All participants major in computer science and have experience in the Python programming language. In preparation for this study, we collected a dataset of incorrect student submissions to programming problems assigned in CS1, an introductory computer science course at our university.

At the start of each study session, we gave each participant a 6-minute tutorial on both Python Tutor and \tool{} to familiarize them with each interface.
We then gave each participant four incorrect submissions (two for \tool{} and two for Python Tutor) and asked to perform two tasks for each problem: (1) point out the location of the bug and (2) fix the bug.
For each incorrect submission, we explained the programming assignment to the participant and then gave them ten minutes to perform the tasks. 
\changes{
Some of the incorrect submissions had multiple bugs to fix, but we did not mention how many bugs the code has.
}
We did not provide means for the participant to \changes{run the program or} check if they had successfully fixed the code, so multiple attempts to fix the code were not allowed.
Once the participants were satisfied with their fix, we ran their code against the test suite to check if the fix had corrected the code. 
After the session, we asked each participant to rate and explain their experience using each interface. We asked participants to evaluate the usefulness of the interfaces along four different aspects: (1) locating the bug, (2) understanding the bug, (3) fixing the bug, (4) learning debugging skills. 
Finally, we conducted a post-survey where participants could compare the experience of using both tools.  

\subsection{Task}
Participants were given incorrect submissions from the following three CS1 programming problems:
\begin{itemize}
\item {\bf Product:} takes as parameters a positive integer $n$ and a unary function $term$, and returns the product of the first n terms in a sequence: $term(1) * term(2) * \cdots * term(n)$.
\item {\bf Accumulate:} takes as parameters the same $n$ and $term$ function as Product as well as a binary function $combiner$ for accumulating terms, and an initial value base. For example, $accumulate(add, 11, 3, square)$ returns $11 + square(1) + square(2) + square(3)$.
\item {\bf Repeated:} takes as parameters a unary function $f$ and a number $n$, and returns the $n$th application of $f$. For example, $repeated(square,2)(5)$ returns $square(square(5))$ evaluates to 625.
\end{itemize}
Our dataset contained 497, 576, and 579 incorrect student submissions for the $product$, $accumulate$, and $repeated$ problems, respectively.

For each problem, we selected two incorrect submissions that contain representative mistakes. To make this selection, we first clustered student mistakes into clusters that share a similar error or mistake. 
We adopted the clustering algorithm from previous work~\cite{head2017writing}. 
We then sorted the clusters by the number of student submissions they contained to identify the most popular mistakes.
We systematically chose the first two clusters as representative mistakes. 
If the second cluster shares the similar fix with the first one, we then skip the second one and pick the next one until the final two clusters have the different mistakes. 
Once we identified the two clusters, we chose one incorrect submission from each cluster. 

Through the pilot study session, we found there is a significant learning effect when the participant works on the same problem. We minimized this effect by using a different programming problem for each condition. 
We also observed differences in problem difficulty. 
For example, some participants found it more difficult to work on the $repeated$ problem compared to the $accumulate$ problem. 
Therefore, we also shuffled the problems for each condition.

\subsection{Result}

\begin{center}
\centering
Table 1. Summary of Study Results \\
(score is average (SD) score of 7-point Likert scale)
\begin{tabular}{ | l | c | c | }
\hline
 & \tool{} & Python Tutor \\
\hline
Correctly identify bugs  & 82.4\% & 76.5\% \\
Fixed bugs      & 70.1\% & 61.8\% \\
Average time to fix the bug & 5:21 (2:49) & 5:03 (2:42) \\
Overall usefulness         & 6.4 (0.9) & 4.7 (2.0) \\
Help to identify the bug   & 5.9 (1.2) & 4.8 (1.7) \\
Help to understand the bug & 5.7 (1.6) & 4.6 (2.0) \\
Help to fix the bug        & 5.3 (2.0) & 3.9 (2.4) \\
Improve debugging skills   & 4.9 (1.9) & 4.7 (1.9) \\
\hline
\end{tabular}
\label{results}
\end{center}



\textbf{RQ1:} Participants using \tool{} correctly identified 82.4\% of the bugs and fixed 70.1\% of all attempts;
participants using Python Tutor correctly identified 76.5\% of the bugs and fixed 61.8\% of them.    
These differences are not statistically significant by the Chi-square test (identify the bug: $\chi^2$ = 0.26, DF = 1, p > 0.6; fix the bug: $\chi^2$ = 0.09, DF = 1, p > 0.7).

\textbf{RQ2:} We did not observe a significant difference in time spent to fix the bug (Average time \tool{}: 5m 21s, Python Tutor: 5m 03s) by the Wilcoxon signed-ranked test (Z = 0.11, p > 0.9).
Although we expected that participants would take less time using \tool{}, we observed that many participants interacted with Python Tutor to make sure they correctly identified the bug using \tool{}. 

\textbf{RQ3:} Although no significant differences were found in the quantitative measures, qualitative feedback from the participants suggests the potential advantages of \tool{} over Python Tutor interface.
64.7\% of the participants believed that \tool{} was more valuable for identifying and fixing bugs and 29.4\% thought that both tools were equally important (only 5.9\% preferred Python Tutor). One participant (P1) mentioned that when using \tool{} he just needed check when the variables had different values to identify the problem in the code.
\changes{
In response to 7-point Likert scale questions, 
participants significantly preferred \tool{} over Python Tutor in four out of five dimensions with Wilcoxon signed-ranked tests: overall usefulness (Z=-2.7, p<0.05), usefulness to identify (Z=-2.6, p<0.05), understand (Z=-2.2, p<0.05), and fix the bug (Z=-2.3, p<0.05). 
}
Most of the participants liked the features of \tool{} (variable filter: 5.9, SD=1.3; comparison of expected behavior: 6.0, SD=1.5) when asked if each feature helped to focus their attention and understand the incorrect behavior. 
Finally, all participants strongly agreed they would like to use \tool{} in programming classes (6.8, SD=0.5).

%% file: 7-discussion.tex
\section{Discussion}
\changes{
Although the participants significantly prefered \tool{} (RQ3), we did not find significant differences in quantitative measures of debugging performance (RQ1 and RQ2).
Based on the observation and the participants' feedback, we would like to discuss several possible reasons for this.
First, both PythonTutor and TraceDiff presume a certain minimum level of knowledge about the Python language.
Students who did not meet that minimum knowledge level were unable to complete some debugging tasks regardless of tool used. 

Second, some of the selected bugs may have been too simple for the students, which may allow them to fix those bugs regardless of tool used. 
Although we tried to select representative and different types of bugs, creating realistic debugging tasks for user studies is a common difficult challenge~\cite{burg2013interactive}.
On the other hand, for one of the incorrect submissions, only participants using \tool{} were able to fix it (5 out of 9), while none of the 8 participants using Python Tutor could fix it.
The correct fix of this submission was: 
}
\begin{lstlisting}
- return repeated(compose1(g, f), n-1)
+ return compose1(repeated(g, n-1), f)    
\end{lstlisting}
\changes{
Participants particularly appreciated the comparison feature of \tool{}, which can illustrate the difference in the control flow and give the idea of swapping the two calls, while this mistake generates a seemingly correct control flow on Python Tutor, which makes it difficult to identify where the error comes from. 
We believe our tool can be more efficient when fixing such complex mistakes.

Finally, since TraceDiff only visualizes a single point where incorrect code diverges from correct code, some students quickly guessed at both the location and extent of the bug and submitted a fix that was only partially complete. 
In our study, this was the student's first and only shot at fixing the code as we did not allow multiple attempts. 
However, if they are allowed to iteratively debug their bugs, they could have received feedback from TraceDiff during the next iteration, which would help them complete their fix.
Therefore, as future work, we are interested in deploying \tool{} to an actual programming course to evaluate the effectiveness of our tool in more realistic situation.
}

%% file: 8-acknowledgements.tex
\changes{
\section*{Acknowledgments}
This research was supported by the NSF Expeditions in Computing award CCF 1138996, NSF CAREER award IIS 1149799, CAPES 8114/15-3, an NDSEG fellowship, a Google CS Capacity Award, and the Nakajima Foundation.
}

%% file: main.bbl
\begin{thebibliography}{10}
\providecommand{\url}[1]{#1}
\csname url@samestyle\endcsname
\providecommand{\newblock}{\relax}
\providecommand{\bibinfo}[2]{#2}
\providecommand{\BIBentrySTDinterwordspacing}{\spaceskip=0pt\relax}
\providecommand{\BIBentryALTinterwordstretchfactor}{4}
\providecommand{\BIBentryALTinterwordspacing}{\spaceskip=\fontdimen2\font plus
\BIBentryALTinterwordstretchfactor\fontdimen3\font minus
  \fontdimen4\font\relax}
\providecommand{\BIBforeignlanguage}[2]{{%
\expandafter\ifx\csname l@#1\endcsname\relax
\typeout{** WARNING: IEEEtran.bst: No hyphenation pattern has been}%
\typeout{** loaded for the language `#1'. Using the pattern for}%
\typeout{** the default language instead.}%
\else
\language=\csname l@#1\endcsname
\fi
#2}}
\providecommand{\BIBdecl}{\relax}
\BIBdecl

\bibitem{corbett2001locus}
A.~T. Corbett and J.~R. Anderson, ``Locus of feedback control in computer-based
  tutoring: Impact on learning rate, achievement and attitudes,'' in
  \emph{Proceedings of CHI}, 2001, pp. 245--252.

\bibitem{guo2015codeopticon}
P.~J. Guo, ``Codeopticon: Real-time, one-to-many human tutoring for computer
  programming,'' in \emph{Proceedings of UIST}, 2015, pp. 599--608.

\bibitem{d2015can}
L.~D'Antoni, D.~Kini, R.~Alur, S.~Gulwani, M.~Viswanathan, and B.~Hartmann,
  ``How can automatic feedback help students construct automata?'' \emph{ACM
  TOCHI}, vol.~22, no.~2, pp. 9:1--9:24, 2015.

\bibitem{glassman2015overcode}
E.~L. Glassman, J.~Scott, R.~Singh, P.~J. Guo, and R.~C. Miller, ``Overcode:
  Visualizing variation in student solutions to programming problems at
  scale,'' \emph{ACM TOCHI}, vol.~22, no.~2, pp. 7:1--7:35, 2015.

\bibitem{head2017writing}
A.~Head, E.~Glassman, G.~Soares, R.~Suzuki, L.~Figueredo, L.~D'Antoni, and
  B.~Hartmann, ``Writing reusable code feedback at scale with mixed-initiative
  program synthesis,'' in \emph{Proceedings of L@S}, 2017.

\bibitem{kaleeswaran2016semi}
S.~Kaleeswaran, A.~Santhiar, A.~Kanade, and S.~Gulwani, ``Semi-supervised
  verified feedback generation,'' in \emph{Proceedings of FSE}, 2016, pp.
  739--750.

\bibitem{rivers2015data}
K.~Rivers and K.~R. Koedinger, ``Data-driven hint generation in vast solution
  spaces: a self-improving {Python} programming tutor,'' \emph{International
  Journal of Artificial Intelligence in Education}, pp. 1--28, 2015.

\bibitem{rolim2017learning}
R.~Rolim, G.~Soares, L.~D'Antoni, O.~Polozov, S.~Gulwani, R.~Gheyi, R.~Suzuki,
  and B.~Hartmann, ``Learning syntactic program transformations from
  examples,'' in \emph{Proceedings of ICSE}, 2017.

\bibitem{singh2013automated}
R.~Singh, S.~Gulwani, and A.~Solar-Lezama, ``Automated feedback generation for
  introductory programming assignments,'' in \emph{Proceedings of PLDI}, 2013,
  pp. 15--26.

\bibitem{shute2008focus}
V.~J. Shute, ``Focus on formative feedback,'' \emph{Review of Educational
  Research}, vol.~78, no.~1, pp. 153--189, 2008.

\bibitem{vanlehn2006behavior}
K.~{VanLehn}, ``The behavior of tutoring systems,'' \emph{International Journal
  of Artificial Intelligence in Education}, vol.~16, no.~3, pp. 227--265, 2006.

\bibitem{suzuki2017exploring}
R.~Suzuki, G.~Soares, E.~Glassman, A.~Head, L.~D'Antoni, and B.~Hartmann,
  ``Exploring the design space of automatically synthesized hints for
  introductory programming assignments,'' in \emph{Proceedings of CHI Extended
  Abstracts}, 2017, to appear.

\bibitem{guo2013online}
P.~J. Guo, ``Online python tutor: Embeddable web-based program visualization
  for {CS} education,'' in \emph{Proceeding of SIGCSE}, 2013, pp. 579--584.

\bibitem{Vanlehn2005}
K.~{VanLehn}, C.~Lynch, K.~Schulze, J.~A. Shapiro, R.~Shelby, L.~Taylor,
  D.~Treacy, A.~Weinstein, and M.~Wintersgill, ``The {Andes} physics tutoring
  system: Lessons learned,'' \emph{International Journal of Artificial
  Intelligence in Education}, vol.~15, no.~3, pp. 147--204, Aug. 2005.

\bibitem{du1986some}
B.~Du~Boulay, ``Some difficulties of learning to program,'' \emph{Journal of
  Educational Computing Research}, vol.~2, no.~1, pp. 57--73, 1986.

\bibitem{ragonis2005understanding}
N.~Ragonis and M.~Ben-Ari, ``On understanding the statics and dynamics of
  object-oriented programs,'' in \emph{ACM SIGCSE Bulletin}, vol.~37, no.~1,
  2005, pp. 226--230.

\bibitem{sorva2012visual}
J.~Sorva \emph{et~al.}, \emph{Visual program simulation in introductory
  programming education}, 2012, {PhD} thesis.

\bibitem{hundhausen2002meta}
C.~D. Hundhausen, S.~A. Douglas, and J.~T. Stasko, ``A meta-study of algorithm
  visualization effectiveness,'' \emph{Journal of Visual Languages \&
  Computing}, vol.~13, no.~3, pp. 259--290, 2002.

\bibitem{sorva2013review}
J.~Sorva, V.~Karavirta, and L.~Malmi, ``A review of generic program
  visualization systems for introductory programming education,'' \emph{ACM
  TOCE}, vol.~13, no.~4, p.~15, 2013.

\bibitem{urquiza2004survey}
J.~Urquiza-Fuentes and J.~A. Vel{\'a}zquez-Iturbide, ``A survey of program
  visualizations for the functional paradigm,'' in \emph{Proceedings of the 3rd
  Program Visualization Workshop}, 2004, pp. 2--9.

\bibitem{ko2008debugging}
A.~Ko and B.~Myers, ``Debugging reinvented: Asking and answering why and why
  not questions about program behavior,'' in \emph{Proceedings of ICSE}, 2008,
  pp. 301--310.

\bibitem{lieber2014addressing}
T.~Lieber, J.~R. Brandt, and R.~C. Miller, ``Addressing misconceptions about
  code with always-on programming visualizations,'' in \emph{Proceedings of
  CHI}, 2014, pp. 2481--2490.

\bibitem{burg2013interactive}
B.~Burg, R.~Bailey, A.~J. Ko, and M.~D. Ernst, ``Interactive record/replay for
  web application debugging,'' in \emph{Proceedings of the 26th annual ACM
  symposium on User interface software and technology}.\hskip 1em plus 0.5em
  minus 0.4em\relax ACM, 2013, pp. 473--484.

\bibitem{burg2015explaining}
B.~Burg, A.~J. Ko, and M.~D. Ernst, ``Explaining visual changes in web
  interfaces,'' in \emph{Proceedings of the 28th Annual ACM Symposium on User
  Interface Software \& Technology}.\hskip 1em plus 0.5em minus 0.4em\relax
  ACM, 2015, pp. 259--268.

\bibitem{oney2009firecrystal}
S.~Oney and B.~Myers, ``Firecrystal: Understanding interactive behaviors in
  dynamic web pages,'' in \emph{Visual Languages and Human-Centric Computing,
  2009. VL/HCC 2009. IEEE Symposium on}.\hskip 1em plus 0.5em minus 0.4em\relax
  IEEE, 2009, pp. 105--108.

\bibitem{ko2004six}
A.~J. Ko, B.~A. Myers, and H.~H. Aung, ``Six learning barriers in end-user
  programming systems,'' in \emph{Proceedings of VL/HCC}, 2004, pp. 199--206.

\bibitem{victor2012learnable}
B.~Victor, ``Learnable programming,'' \emph{Worrydream. com}, 2012.

\bibitem{victor2011up}
------, ``Up and down the ladder of abstraction,'' \emph{Retrieved September},
  vol.~2, p. 2015, 2011.

\bibitem{pdb}
\BIBentryALTinterwordspacing
``\texttt{pdb}---the python debugger.'' [Online]. Available:
  \url{https://docs.python.org/2/library/pdb.html}
\BIBentrySTDinterwordspacing

\end{thebibliography}
